\documentclass[english,conference,letter]{IEEEtran}
\usepackage[LGR,T1]{fontenc}
\usepackage[latin9]{inputenc}
\usepackage{geometry}
\usepackage{color}
\usepackage{amsmath}
\usepackage{amssymb}
\usepackage{graphicx}
\usepackage{algorithmic}
\usepackage{algorithm}
\usepackage{subfigure}
\makeatletter

\DeclareRobustCommand{\greektext}{%
  \fontencoding{LGR}\selectfont\def\encodingdefault{LGR}}
\DeclareRobustCommand{\textgreek}[1]{\leavevmode{\greektext #1}}
\DeclareFontEncoding{LGR}{}{}
\DeclareTextSymbol{\~}{LGR}{126}

\usepackage{babel}

\makeatother

\usepackage{babel}

\IEEEoverridecommandlockouts

\allowdisplaybreaks

\title{LP formulations for secrecy\\ over erasure networks with feedback}
\author{
\authorblockN{Athanasios Papadopoulos}
\authorblockA{
UCLA, Los Angeles, USA\\
\textsf{athanasios.papadopoulos@ucla.edu}}
\and
\authorblockN{Laszlo Czap}
\authorblockA{
EPFL, Switzerland\\
\textsf{laszlo.czap@epfl.ch}}
\and
\authorblockN{Christina Fragouli}
\authorblockA{
UCLA, Los Angeles, USA \\
\textsf{christina.fragouli@ucla.edu}}
}
\begin{document}
\maketitle
\begin{abstract} 
We design polynomial time  schemes for secure message transmission over arbitrary networks, in the presence of an eavesdropper, and where each edge corresponds to an erasure channel with public feedback. Our schemes are described through  linear programming (LP) formulations, that explicitly select (possibly different) sets of  paths for key-generation and message sending.   Although our LPs are not always capacity-achieving, they outperform the best known alternatives in the literature, and extend to incorporate several interesting scenaria.
\end{abstract}

\textcolor{white}{.}

\section{Introduction}
We consider the following setup.
 A source, Alice, is connected to a destination, Bob, over a packet network that can be represented as an arbitrary directed acyclic graph. Alice wants to send a message to Bob, securely from a passive eavesdropper, Eve, who wiretaps an unknown subset of $k$ edges in the network.  Each edge $i$ that connects node $u$ to node $v$ corresponds to a packet erasure channel with probability $\delta_i$; when eavesdropping this edge, Eve   also receives the packet transmissions of node $u$ with erasure probability $\delta_{iE}$, independently from node $v$. Moreover, we assume that all legitimate nodes in the network, as well as Eve, causally learn whether $v$ has successfully received the packets $u$ transmits or not; 
however, Eve does not report which packets she successfully receives.

We propose the first, as far as we know, linear programming (LP) formulation, that  explicitly selects paths
in the network to maximize the secure message transmission rate.
 It is well known that the (non-secure) capacity of a network can be described  by an LP which allows
a natural flow-based interpretation of network traffic. Our work leverages this formulation to implement
secure message transmission  through a two-phase construction.  In the first {\em key-creation} phase,  Alice  establishes a secret key 
with Bob; in the second {\em message-sending} phase, she uses the established secret key to encode and securely send a message. Accordingly, our LP selects two sets of paths (that share the network resources): key-creation paths, that Alice will use to share random packets with Bob (so as to  create a secret key), and message-sending paths, that Alice will use to send the encrypted message. We term this end-to-end encryption algorithm (Algo 1). We discuss several extensions of Algo 1, notably Algo 2, that apart from the end-to-end key, also creates and utilizes link-by-link keys for secure message transmission.

The LPs we propose are not optimal,  but are still we believe interesting.
An example where the LPs are suboptimal  is the triangle network, where the capacity was characterized in \cite{triangle14} . However, there are also a number of examples where the LPs do achieve the known capacity (such as the two-parallel edges network, and the line network); they outperform the best known alternative in the literature in all the cases that we have tested; and they enable new observations. For instance, over erasure networks, there are cases where the key-sharing and message-sending paths use different edges  (while over lossless networks, using the same sets of paths is optimal).

Another  attractive attribute of the proposed LPs is  their generality: the LPs take as input
the erasure probabilities $\delta_i$ and $\delta_{iE}$ at every channel edge $i$,
that can be arbitrary. For instance, with  $\delta_i=\delta_{iE}=0$  we recover the lossless
network case, and the LPs achieve the secure network coding rate (which is the optimal
scheme for lossless channels).  
Moreover, similarly to the max-flow LP,  our LPs can be extended to incorporate multiple sources, multiple receivers, edges with costs, etc.

The paper is organized as follows. Section~\ref{sec:related} briefly reviews related work; Section~\ref{sec:not}
introduces our notation; Section~\ref{sec:algo}  presents  the algorithms; and Section~\ref{sec:eval} has evaluation results.

\section{Related Work} \label{sec:related}
Finding the highest achievable rate of secure communication 
of an arbitrary network setting is an open research problem. In the special case when the network
consists of {\em error-free}, unit capacity channels, secure network coding is optimal \cite{cai2011}. 
For the same problem when the channels are not unit capacity (but still error-free)
restricted complexity results suggest the hardness of calculating the secret message capacity
\cite{cui2010,cui2013}. When the network edges are erasure channels {\em all
with the same parameters} and channel state feedback, and  the paths used for Alice to
communicate with Bob are decided in advance, a secure communication achievable scheme is proposed
in \cite{Allerton2013}. In contrast, this work provides schemes for arbitrary erasure channel
parameters, and explicitly selects the best paths in the network so as to maximize the achievable
rates. For a number of small networks (single channel, V-network, triangle network, line network)
 with erasures and state feedback, capacity characterization and
a linear programming formulation were derived in \cite{triangle14,JDFPA10,ITW11,isit12,itw13,athan2014}. Our approach in this work is different: instead
of schemes tailored to specific topologies, we design schemes that are general and extend to
arbitrary network topologies.
A preliminary version of  LP formulations (a precursor of the algorithm we call Algo 1) for this problem was presented as an invited poster in a workshop \cite{GlobalSip}. 

\newpage
\section{\label{sec:not}System Model and Notation}

A source $s$ (Alice)  wants to send a message $W$ securely to a destination $d$ (Bob), over a
 directed acyclic graph $G=(\mathcal{V},\; \mathcal{E})$, where each edge $g$ that connects node $u$ to node $v$ 
represents an orthogonal discrete memoryless broadcast 
erasure channel with two receivers: node $v$ and potentially a passive eavesdropper (Eve).
We denote by $X_{gi}$  the input to channel $g$ at time slot $i=1\ldots n$;   
and by $Y_{gi}$ and $Z_{gi}$ the corresponding outputs  at node $v$ and Eve respectively. 
We assume that  $X_{gi}$  is a length
$L$ vector over a finite field $\mathbb{F}_{q}$
 (in the paper we use the convention that
$L\log(q)=1$). 
We use $\oslash$ as the symbol
of an erasure.
The broadcast channel is conditionally independent, namely
\[ 
\Pr\{Y_{gi}^{n},Z_{gi}^{n}|X_{gi}^{n}\}=\underset{i=1}{\overset{n}{\prod}}\Pr\{Y_{gi}|X_{gi}\}\Pr\{Z_{gi}|X_{gi}\},
\]
\begin{align*}
\mbox {with} \quad\quad\quad\quad
\Pr\{Y_{gi}|X_{gi}\} & =\begin{cases}
1-\delta_{g}, & Y_{gi}=X_{gi}\\
\delta_{g}, & Y_{gi}=\oslash,
\end{cases}
\end{align*}
\begin{eqnarray*}\mbox{and}\quad\quad 
\Pr\{Z_{gi}|X_{gi}\} & =\begin{cases}
1-\delta_{gE}, & Z_{gi}=X_{gi}\\
\delta_{gE}, & Z_{gi}=\oslash.
\end{cases}
\end{eqnarray*}
We assume that the source has unlimited private randomness, and that all other network nodes have no private randomness.
 We also assume public
state feedback, that is, each  legitimate node sends an ACK (or NACK) so that all other nodes (including
Eve) learn whether the packet transmission was successful. 
We use
the notation $S^{i-1}$ for the state of all the channels before the transmission of the $i^{th}$
symbols. Also the notation
$I_{u}$ and $O_{u}$ for the set of the incoming and outgoing edges of node~$u$.

We require security
 in the strong information theoretical sense, defined next  in the same way as in \cite{Allerton2013,ITW11}.
 We use $X_{Ai}$, for a set A, to  denote the vector $(X_{gi})_{g\in A}$, and
 $X_{A}^{i}$  to denote the vector $(X_{A1},X_{A2},\dots,X_{Ai})$.
%
\\
 {\bf Definition.} We say that $R_{SM}$ is an achievable secret message rate if
for any $\epsilon>0$ and sufficiently large $n$ the following conditions hold for some functions
$f_{gi,n}(\cdot),W_{B,n}(\cdot)$.\\
For $u\in U-\{s\}$ and for every $g\in O_{u}$:
\begin{align}
X_{gi}=\begin{cases}
f_{gi,n}(Y_{I_{u}}^{i-1},S^{i-1})\end{cases}\label{eq:def1_1}
\end{align}
\begin{align}\mbox{and for every $g\in O_{s}$:} \quad\quad
X_{gi}=\begin{cases}
f_{gi,n}(W,U_{0},S^{i-1})\end{cases}\label{eq:def1_1-1}
\end{align}
where $U_{0}$ is the unlimited random source of Alice and where the message $W$ is uniformly
distributed over $\{1,2,\ldots,2^{n(R_{SM}-\epsilon)}\}$. Bob is able to recover $W$ with
high probability, 
\begin{align}
\hat{W}=W_{B,n}(Y_{I_{d}}^{n}),\label{eq:def1_2}\\
\Pr\{\hat{W}\neq W\}<\epsilon.\label{eq:def1_3}
\end{align}
Eve gains negligible useful information by observing $V\subseteq\mathcal{E}$: 
\begin{align}
I(W;Z_{V}^{n}S^{n})<\epsilon,\:\forall \;\;V\subseteq\mathcal{E}.
\end{align}
The supremum of all achievable secret message rates is the secret message capacity of the network
denoted by $C_{SM}$.

\section{ End-to-End Encryption Algorithm}
\label{sec:algo}

\paragraph*{Broad Approach}
The algorithm selects two (possibly different) sets of paths, one set for key-creation and the other for message-sending.
The source uses the first (key-creation) set of paths to send random packets to the destination; intermediate
nodes forward the random packets they receive from  their incoming edges to their outgoing edges using two techniques, ARQ and MDS expansion, as we will describe later in this section. The source and the destination create an end-to-end secret key, based on their shared random packets and an estimate of how many of these Eve has eavesdropped.
The algorithm also selects a second set of paths, over which the source sends an information message to the destination, encrypted with the source-destination end-to-end key. Intermediate nodes simply forward the encrypted packets using ARQ. The goal of the program is to maximize
the rate at which the message can be send securely to the destination, by optimizing over two things: 1) what are the paths selected for key-creation and message-sending and 2) and how are the random packets forwarded by the intermediate nodes.


\subsection{Scheme Description and Algo 1 LP}
We start from the case where Eve observes any (one) edge of the network.
All  the LP variables express rate of packets, either message-packets, or random-packets.

\paragraph*{Key-creation constraints}
The source generates uniform random packets, to be send to the destination.
The intermediate nodes collect  the random packets they receive from all  their incoming edges,  partition them into subsets, and send a subset to each of their outgoing edges using two techniques, Automatic Repeat Request (ARQ) and Maximum-Distance-Seperable (MDS) code expansion. 
To capture this, for each edge (channel) $g$, that connects say vertex $u$ to vertex $v$,  the LP uses three variables $s_g$, $k_g$ and $r_g$. 
Node $u$ sends $k_g$ packets to node $v$,
by first multiplying these packets with an MDS code of size $\frac{k_{g}}{1-\delta_{g}\delta_{gE}} \times k_g$ to create  $\frac{k_{g}}{1-\delta_{g}\delta_{gE}}$ linear combinations, and then transmitting each linear combination exactly once (we discuss later why we expand with these parameters).   From these packets, $v$ receives a fraction $k_{g}\frac{1-\delta_{g}}{1-\delta_{g}\delta_{gE}}$. 
Moreover, $u$ also sends to node $v$   $r_{g}$ packets using ARQ; $v$ receives all these packets. 
Node $v$ receives in total rate $s_g$ random packets, with 
\begin{align}
s_{g}=r_{g}+k_{g}\frac{1-\delta_{g}}{1-\delta_{g}\delta_{iE}}.
\end{align}
If node $u$ has  $I_{u}$ incoming and $O_{u}$ edges, we have that
\begin{align}
\sum_{i\in I_{u}}s_{i}=\sum_{j\in O_{u}}(k_{j}+r_{j}).
\end{align}
This constraint requires that  the random packets node $u$ sends are equal to the random packets it receives;
that is,  intermediate network nodes do not discard or generate random packets. 

\paragraph*{Message-sending constraints}
The source encrypts the message using an end-to-end key (we will describe how later), and forwards it to the destination; each intermediate node uses ARQ to forward the encrypted message packets it receives.
The LP  uses a variable $m_g$ to capture the encrypted message packets 
that node $u$ sends to node $v$ through the edge $g$ that connects them; note that to do so,  
node $u$ makes $\frac{m_{g}}{1-\delta_{g}}$ transmissions.
We require  message flow conservation, i.e.,
\begin{align}
\sum_{i\in I_{u}}m_{i} & =\sum_{j\in O_{u}}m_{j}.
\end{align}

\paragraph*{Time-sharing (edge capacity) constraints}
Random and encrypted packets need to potentially share the network edges (channels); we thus require for every edge of the network that 
\begin{align}
 \frac{r_{g}}{1-\delta_{g}}+\frac{k_{g}}{1-\delta_{g}\delta_{gE}}+\frac{m_{g}}{1-\delta_{g}}\leq 1. 
\end{align}

\paragraph*{Security constraints} If Eve is located on edge $g$, she will overhear a fraction
\begin{align*}
m_{g}\frac{1-\delta_{gE}}{1-\delta_{g}\delta_{gE}}
\end{align*}
of the encrypted message flow $m_g$ through that edge. 
A necessary condition for our scheme to be secure is that,
this amount of message is smaller than the  amount of random packets that Alice and Bob have and Eve does not, i.e., the secret common random packets (this condition is also sufficient for security as we discuss later on).
Alice and Bob share $\left(\sum_{j\in I_{D}}s_{j}\right)$ random packets; thus if, from these 
  $\left(\sum_{j\in I_{D}}s_{j}\right)$ packets, Eve has overhead say $E_g$ (by observing the random packet flow through edge $g$), then the security constraint would be:
\begin{align*}
m_{g}\frac{1-\delta_{gE}}{1-\delta_{g}\delta_{gE}} & \leq\left(\sum_{j\in I_{D}}s_{j}\right)- E_g.
\end{align*}


\paragraph*{Conservatively estimating Eve's knowledge $E_g$} Consider again edge $g$ that connects vertex $u$ to vertex $v$.  A conservative way to estimate Eve's knowledge,
is to set 
\[
E_g=r_{g}\frac{1-\delta_{gE}}{1-\delta_{g}\delta_{gE}} + k_{g}\frac{(1-\delta_{gE})(1-\delta_{g})}{1-\delta_{g}\delta_{gE}}.\]
That is, calculate the number of random packets that both node $v$ and Eve receive.
This estimate is conservative because we assume that all the randomness node $v$ receives  eventually reaches the destination (Bob), which is not necessarily true.  Indeed, when nodes forward packets using the MDS expansion,
we "lose" part of the randomness (from the $k_g$ random packets, node $u$ only receives $k_{g}\frac{(1-\delta_{g})}{1-\delta_{g}\delta_{gE}}).$  Algo 1 uses this approximation.

\renewcommand{\thealgorithm}{}
{\small
\begin{algorithm}
\floatname{algorithm}{Algo 1}
\begin{algorithmic}
\renewcommand{\algorithmicrequire}{\textbf{Input:}}
  \REQUIRE Set of erasure probabilities $\delta_g$ and $\delta_{gE}$
\renewcommand{\algorithmicrequire}{\textbf{Output:} Secure message rate and achievability scheme parameters}
  \REQUIRE 
\begin{align}
\max R,\text{s.t.:}\nonumber\\
R & =\sum_{i\in I_{D}}m_{i}\nonumber\\
\forall u\in \mathcal{V}-\{s,d\}:\nonumber\\
\sum_{i\in I_{u}}m_{i} & =\sum_{j\in O_{u}}m_{j}\nonumber\\
\sum_{i\in I_{u}}s_{i} & =\sum_{j\in O_{u}}(k_{j}+r_{j})\nonumber\\
\forall g\in \mathcal{E}:\nonumber\\
s_{g} & =r_{g}+k_{g}\frac{1-\delta_{g}}{1-\delta_{g}\delta_{gE}}\nonumber\\
1 & \geq\frac{r_{g}}{1-\delta_{g}}+\frac{k_{g}}{1-\delta_{g}\delta_{gE}}+\frac{m_{g}}{1-\delta_{g}}\nonumber\\
m_{g}\frac{1-\delta_{gE}}{1-\delta_{g}\delta_{gE}} & \leq\left(\sum_{j\in I_{D}}s_{j}\right)-r_{g}\frac{1-\delta_{gE}}{1-\delta_{g}\delta_{gE}}\nonumber\\
 & \;\:-k_{g}\frac{(1-\delta_{gE})(1-\delta_{g})}{1-\delta_{g}\delta_{gE}}\nonumber\\
 \forall\; i:\; \;m_i, s_i, k_i, r_i &\geq 0. \nonumber 
\end{align}
\end{algorithmic}
\caption{LP with end-to-end encryption and $E_g$ approximation}
\label{algo1}
\end{algorithm}
}

\paragraph*{Message encryption at the source}
The LP identifies the rate $R$ at which we can send an encrypted message, and the rates $m_g$ of the message that flow through each edge. We encrypt the message using a one-time pad approach and a key of size $R$, that we create by multiplying the $ \sum s_i$ packets that Bob receives with an  $ R \times \sum s_i$ MDS matrix.

\subsection{Discussion}

\paragraph*{Why use MDS expansion at intermediate nodes}
When the network consists of a single edge, the optimal key-generation scheme has Alice generate uniform at random packets and send these to Bob \cite{ITW11}; this has the advantage that packets that Eve receives and Bob does not, give no information to Eve about the packets Bob receives. Using MDS at intermediate nodes mimics this effect more efficiently: the observation is that, if Alice sends uniform at random packets, there exist some packets that neither Bob nor Eve receive; thus in a sense these packets do not serve any purpose. To avoid this, Alice can simply expand the $k$ packets to $\frac{k}{1-\delta\delta_E}$ packets. MDS  combining has the property that Eve cannot learn anything about the packets that Bob receives, from the packets that only she (and not Bob) has collected. This observation and the corresponding proof were provided in \cite{athan2014}. The LP selects what fraction of the packets to send using MDS, and what fraction to send using ARQ, separately for each edge. ARQ  has the advantage that it preserves all random packets, and the disadvantage that Eve learns more about the packets that Bob collects.

\paragraph*{Why ARQ for message sending}
ARQ is a capacity achieving strategy over erasure channels, as is also erasure coding. However, when we are interested in secure message sending,  if we were to take the message, encrypt it with a one-time pad, and then use erasure correcting coding to transmit it to Bob, we would get a worse performance than if we send the encrypted message with ARQ. This is beacuse, with erasure coding, {\em every} packet Eve receives gives her {\em new} information about the information message; however, with ARQ, she may receive repeated packets, that bring her no new information.

\paragraph*{Exact calculation of $E_g$}
One method is similar to the standard path-LP formulation of the (non-secure) max flow LP, i.e., the LP that assigns rates to each of the paths that connect a source to a destination.  To calculate $E_g$, we associate with each path $p$ a random packet flow $s_p$ that captures the delivered random packets through that path from Alice to Bob. We can then calculate how many of  the packets Bob receives are delivered through paths that include edge $g$, and remove the fraction that Eve overhears.     
This approach has a compact LP form and is  illustrated in Algo 2. Although this formulation has exponential complexity,
it is also possible to exactly calculate $E_g$ in polynomial time (see Appendix).  For this, we need to assume that network nodes do an additional operation:
every node in the network uniformly at random mixes its incoming random packets before forwarding them towards Bob; we thus ensure that "all packets are treated equally". We then reduce the problem to calculating, what fraction of random packets that go through a given node, reach Bob.

\subsection{Analysis}

\paragraph*{Why the scheme is secure} This follows directly by applying Theorems 10 and 11 of \cite{athan2014} as well as Lemma 4 of  \cite{JDFPA10}. For completeness we include  a proof in the Appendix.


\paragraph*{Reduction to secure network coding}
By setting $\delta_i=\delta_{iE}=1$ for every edge of the network, the solution of the Algo 1 LP gives the same result as 
secure network coding. Indeed, if we assume that the mincut equals $h$, selecting $h$ edge-disjoint paths, and using
$h-1$ of them to end the encrypted message and one  to send random packets for key generation, is a feasible
solution. From \cite{cai2011} it is also an optimal solution for this network.

\paragraph*{Suboptimality} The achievability algorithm we presented is suboptimal, not only because it uses an estimate for $E_i$, but also because it only creates an end-to-end key; we know from the work in \cite{triangle14}, that, to achieve the capacity in some cases, even  when the intermediate nodes do not have  private randomness, we need to create and explore common randomness they have by receiving the same source-generated random packets, leading to an exponential complexity problem \cite{cui2010,cui2013}. 

\paragraph*{Optimality in some cases} In some cases where the secure message capacity is known,
we can prove that Algo 1 (or Algo 2 we describe later) are optimal. For illustration, we provide in the Appendix a proof that Algo 1 is optimal when Alice is connected to Bob through two parallel channels. Algo 2 achieves the capacity of the line network, as again shown in the Appendix.

\subsection{Extensions}\label{sec:ext}
Given the framework of Algo 1, we
can directly extend it in a number of cases, as is also the case for the max flow LP, albeit at  additional complexity cost in some cases.
For instance, we can extend it to address the following:\\
$1.$ Link-by-link key creation (see for example  Algo 2).\\
$2.$ Multicasting to more than one receivers  (following a similar approach to the network coding LP in \cite{Allerton2013,Li2006}).\\
$3.$ Eve wiretaps more than one edges  (if Eve eavesdrops $V$ edges, $E_g$ would be the amount of random packets Eve has collected by eavesdropping on edge $g$ plus $V-1$ arbitrary other edges.  We provide such an LP in the  Appendix for illustration.)\\
$4.$ Multiple sources not collocated transmitting messages to the same receiver (in this case, we can combine random packets across sources to create link by link keys; see Appendix).\\
 $5$. Having costs associated with edges (similarly to \cite{athan2014}).
\renewcommand{\thealgorithm}{}
{\small
\begin{algorithm}
\floatname{algorithm}{Algo 2}
\begin{algorithmic}
\renewcommand{\algorithmicrequire}{\textbf{Input:}}
  \REQUIRE Set of erasure probabilities $\delta_g$ and $\delta_{gE}$
\renewcommand{\algorithmicrequire}{\textbf{Output:} Secure message rate and achievability scheme parameters}
  \REQUIRE 
\begin{align}
\max R,\text{s.t.:}\nonumber\\
R & =\sum_{i\in I_{D}}m_{i}\nonumber\\
\forall u\in \mathcal{V}-\{s,d\}:\nonumber\\
\sum_{i\in I_{u}}m_{i} & =\sum_{j\in O_{u}}m_{j}\nonumber\\
\sum_{i\in I_{u}}s_{i} & \geq\sum_{j\in O_{u}}(k_{j}+r_{j})\nonumber\\
\forall g\in \mathcal{E}:\nonumber\\
s_{g} & =r_{g}+k_{g}\frac{1-\delta_{g}}{1-\delta_{g}\delta_{gE}}\nonumber\\
1 & \geq\frac{r_{g}}{1-\delta_{g}}+\frac{k_{g}}{1-\delta_{g}\delta_{gE}}+\frac{m_{g}}{1-\delta_{g}}\label{eq:noc}\\
s_{g} & =\sum_{p\in P:g\in p}s_{p} \label{eq:sum}\\
m_{g}\frac{1-\delta_{gE}}{1-\delta_{g}\delta_{gE}} & \leq(k_{g}+r_{g})\frac{\delta_{gE}(1-\delta_{g})}{1-\delta_{g}\delta_{gE}}+\sum_{p\in P_{-g}^{'}}s_{p}
\nonumber\\
 \forall\; i,p:\; \;m_i, s_i, s_p, k_i, r_i &\geq 0. \nonumber 
\end{align}
\end{algorithmic}
\caption{LP with end-to-end and link-by-link encryption, and with $E_g$ exact calculation}
\label{algo2}
\end{algorithm}
}

\paragraph*{Algo 2 description}
In this algorithm the message is encrypted both with an end-to-end key, and a link-by-link
key (that is applied and peeled off at every edge). The source again selects two (possibly
different) sets of paths, one set for random-packet-sending and the other for message-sending.
An end-to-end key is created from these random packets. The source encrypts all the packets
with this end-to-end key and transmits them appropriately through the message-sending paths. 

Furthermore, node $u$ (connected to node $v$ through edge $g$) may also apply an additional
link-by-link key, that node $v$ will remove before further forwarding and potentially re-encrypting
the message. Note that $u$ may send to node $v$ more random packets than what node
$v$ can forward to Bob, as these extra packets are still useful to create a larger link-by-link
key for edge $g$. Algo 2 uses all the $s_{g}$ random packets to create the link-by-link key.
These packets can no longer contribute to the end-to-end key that will also protect the message
$m_{g}$ through edge $g$, and need to be accounted for. 

Algo 2 exactly calculates how many of the $s_{g}$ packets reach Bob, through a path flow-decomposition
approach. We denote with $P$ the set of all paths in the network that begin from the source,
with $P'$ all the Alice-Bob paths, and with $P_{-g}^{'}$ all Alice-Bob paths that do not
utilize edge $g$. The LP assigns values to each message-path-flow $s_{p}$ and of course it
is,

\[
s_{g}=\sum_{p\in P:g\in p}s_{p}.
\]

For the calculation of the key for edge $g$: 

The link-by-link key is calculated as the random packets that pass through edge $g$ and are
not heard by Eve,

\[
(k_{g}+r_{g})\frac{\delta_{gE}(1-\delta_{g})}{1-\delta_{g}\delta_{gE}}.
\]

The end-by-end key is calculated as the random packets that were transmitted to the destination
without passing through edge $g$,

\[
\sum_{p\in P_{-g}^{'}}s_{p}.
\]

Since we are protecting from an Eve at edge $g$, we are sure that all these packets are secure.

Thus the security condition becomes,

\[
m_{g}\frac{1-\delta_{gE}}{1-\delta_{g}\delta_{gE}}\leq(k_{g}+r_{g})\frac{\delta_{gE}(1-\delta_{g})}{1-\delta_{g}\delta_{gE}}+\sum_{p\in P_{-g}^{'}}s_{p}.
\]

The LP can choose among many different path-flows for the messages in order to achieve the
same $s_{g}$ for all edges $g$. The optimal choice is the one that
maximizes the secure message sending rate.

\begin{figure*}[t!]
\centering
\subfigure[Message-sending and key-creation paths can be different: the upper path is used only for message flow, the lower path is shared. We depict the optimal values Algo I has selected. ]
  {\includegraphics[width=0.8\columnwidth]{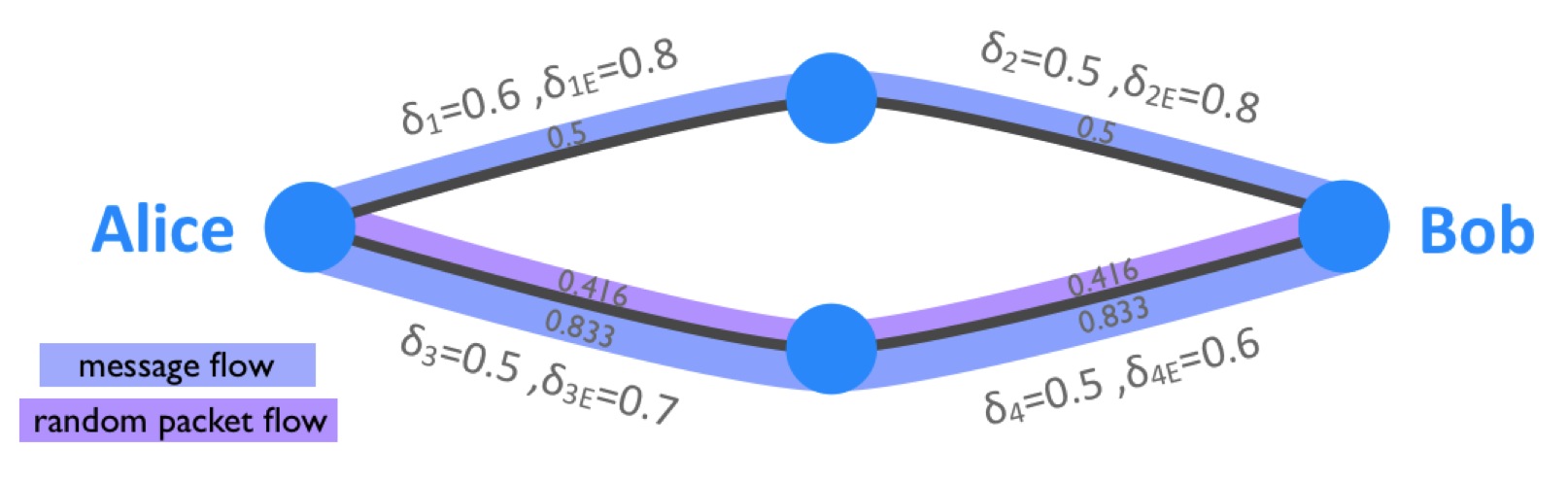}
\label{fig1}
}
\subfigure[Line network with $N+1$ nodes.]{
  \includegraphics[width=0.8\columnwidth]{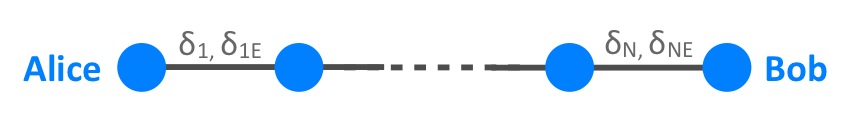}
\label{fig:Line}
}
\caption{Network configurations.}
\label{fig:2rel_networks}
\end{figure*}

\begin{figure*}[t!]
\centering
\subfigure[Two hop line network
with  $\delta_{\text{2\textgreek{E}}}=\delta_{\text{1\textgreek{E}}}=\delta_{\text{\textgreek{E}}}$, $\delta_{\text{1}}=0.2$, $\delta_{\text{2}}=0.8$.]
  {\includegraphics[width=0.65\columnwidth]{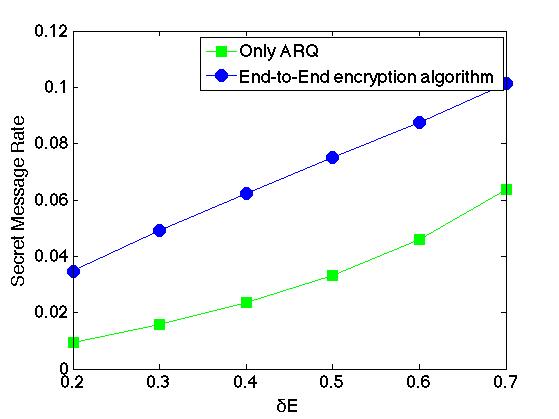}
\label{fig3}
}
\subfigure[Multiple parallel channels with $\delta_{\text{i\textgreek{E}}}=0.8$,
$\delta_{\text{i}}=0.6$ for $i$  odd, and $\delta_{\text{i\textgreek{E}}}=0.9$, $\delta_{\text{i}}=0.6$
for $i$  even.] 
{
  \includegraphics[width=0.65\columnwidth]{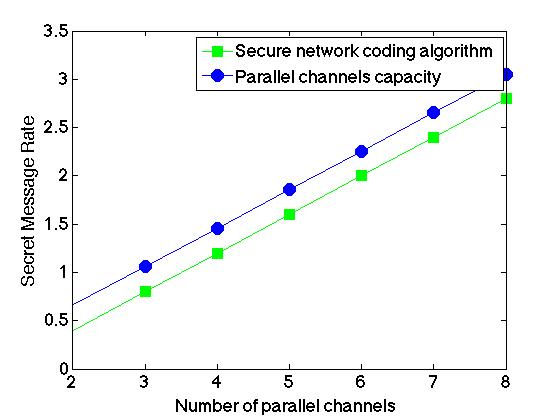}
\label{fig4}
}
\subfigure[Two hop line network 
with  $\delta_{\text{1\textgreek{E}}}=0.5$,  
$\delta_{\text{2\textgreek{E}}}=1$, $\delta_{\text{2}}=0.6$.]
{\includegraphics[width=0.65\columnwidth]{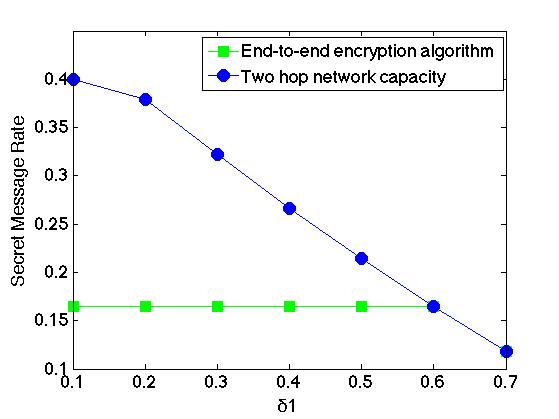}
\label{fig5}
}
\caption{Evaluation results through matlab.} 
\label{fig:2rel_networks}
\end{figure*}

\newpage
\section{Evaluation}\label{sec:eval}
We used numerical evaluations (through matlab) to solve the LPs  over specific configurations where the capacity is known.
We verified that: \\
$\bullet$ \emph{Selecting paths helps.}
The optimal message-sending and key-creation sets of paths in several instances did not share all edges.
Such an example  is provided in Fig.~\ref{fig1}.\\
$\bullet$ \emph{Generating keys using MDS helps.}
Fig.~\ref{fig3} shows the performance we get over a two-hop line network (Fig.~\ref{fig:Line} with $N=2$), when: 1) we allow the LP in Algo~1 to only use ARQ for the random packets propagation to the destination, and 2) we use both ARQ and MDS for the same purpose. The benefits of using MDS in this case are clear. Note that over the line network secure network coding achieves zero rate.\\
$\bullet$ \emph{Algo 1 is suboptimal,} Fig.~\ref{fig5} compares the performance of Algo 1 with the capacity of the two-hop line network \cite{athan2014}; when Eve only wiretaps the first channel, and the first channel is better than the second, the optimal strategy uses a link-by-link key; Algo 1 cannot do this. Algo 2, that can do so,  achieves the capacity.\\ 
$\bullet$ \emph{Using link-by-link keys can  help.} See  previous point. \\
$\bullet$ \emph{We achieve benefits over secure network coding.} We compare Algo 1 against using channel coding followed by secure network coding. Fig.~\ref{fig4} considers a configuration where Alice is connected to Bob through multiple parallel channels; this is a "worse case" configuration in terms of expected benefits, as the main opportunity to create keys comes from the number of paths (and not erasures), that secure network coding also leverages. The constant benefits Algo 1 offers are exactly due  to  exploiting the erasures over the edge that Eve wiretaps.

\bibliographystyle{IEEEtran}
\bibliography{bibliography}

\newpage
\section{Appendix}

We provide the following  at the interested reviewer's discretion:
A) Optimality of Algo~1 for two parallel channels; B) Optimality  of Algo~2 for line network;
C)  Calculation of $E_g$  in polynomial time;
 D) Security of Algo~1;   and E) Examples of extending the LPs.


\subsection{Optimality of Algo 1 for the two parallel channels network}
The outerbound  in \cite{itw13} for the two parallel channels is:
\begin{align}
\max M,\text{s.t.:}\nonumber\\
M&=(1-\delta_{1})M_{1}+(1-\delta_{2})M_{2}\nonumber\\
1 & \geq C_{1}+M_{1}\nonumber\\
1 & \geq C_{2}+M_{2}\nonumber\\
M_{1}\frac{(1-\delta_{1E})(1-\delta_{1})}{1-\delta_{1}\delta_{1E}} & \leq C_{2}(1-\delta_{2})+C_{1}(1-\delta_{1})\delta_{1E}\nonumber\\
M_{2}\frac{(1-\delta_{2E})(1-\delta_{2})}{1-\delta_{2}\delta_{2E}} & \leq C_{1}(1-\delta_{1})+C_{2}(1-\delta_{2})\delta_{2E}.\nonumber
\end{align}
A feasible solution for the Algo 1 LP  is $r_{1}=0$, $r_{2}=0$. In this case, making the correspondence
$m_{i}\rightarrow(1-\delta_{i})M_{i}$ , $k_{i}/(1-\delta_{i}\delta_{iE})\rightarrow C_{i}$
for $i=\{1,2\}$, we can see that the two LPs are equivalent. Thus the end-to-end encryption
algorithm achieves the capacity of the parallel channels network.

\subsection{Optimality of Algo  2 for line network}

The outerbound derived in \cite{athan2014} for the line network is:
\begin{eqnarray*}
max & m,\\
s.t.\:\ensuremath{\forall j\in\mathcal{N}}:\\
\frac{1-\delta_{jE}}{1-\delta_{j}\delta_{jE}}m & \leq & k_{j}\\
\frac{k_{j}}{(1-\delta_{j})\delta_{jE}}+\frac{m}{1-\delta_{j}} & \leq & 1\\
k_{j} & \leq & d_{j-1}\frac{\delta_{jE}(1-\delta_{j})}{1-\delta_{j}\delta_{jE}}\\
d_{j}+m & \leq & 1-\delta_{j}\\
d_{j} & \leq & d_{j-1},\: j>1
\end{eqnarray*}
In this case there is only one path and Algo 2 becomes equivalent to the outerbound
of the line network, and thus achieves the
capacity of the line network.

{\small
\begin{algorithm}
\floatname{algorithm}{Algo  3}
\begin{algorithmic}
\renewcommand{\algorithmicrequire}{\textbf{Input:}}
 \REQUIRE Set of erasure probabilities $\delta_i$ and $\delta_{iE}$.
\renewcommand{\algorithmicrequire}{\textbf{Output:} Secure message rate and achiev. scheme parameters}
  \REQUIRE 
\begin{align}
\max R,\text{s.t.:}\nonumber\\
R & =\sum_{i\in I_{D}}m_{i}\nonumber\\
\forall u\in \mathcal{V}-\{s,d\}:\\
\sum_{i\in I_{u}}m_{i} & =\sum_{j\in O_{u}}m_{j}\nonumber\\
\sum_{i\in I_{u}}s_{ii} & =\sum_{j\in O_{u}}(k_{j}+r_{j})\nonumber\\
\forall g\in \mathcal{E}:\nonumber\\
s_{gg} & =r_{g}+k_{g}\frac{1-\delta_{g}}{1-\delta_{g}\delta_{gE}}\label{eq:delivered_randomness-4}\nonumber\\
1 & \geq\frac{r_{g}}{1-\delta_{g}}+\frac{k_{g}}{1-\delta_{g}\delta_{gE}}+\frac{m_{g}}{1-\delta_{g}}\nonumber\\
\forall u\in \mathcal{V}-\{s,d\},\nonumber\\
\forall g\in \mathcal{E}:\nonumber\\
\sum_{j\in O_{u}}s_{gj} & \geq\sum_{j\in I_{u}}s_{gj}-\left(\sum_{j\in I_{u}}s_{jj}-\sum_{j\in O_{u}}s_{jj}\right)\nonumber\\
\forall g,j\in \mathcal{E}:\nonumber\\
s_{gj} & \leq s_{jj}\nonumber\\
\forall g\in \mathcal{E}:\nonumber\\
m_{g}\frac{1-\delta_{gE}}{1-\delta_{g}\delta_{gE}} & \leq\left(\sum_{j\in I_{d}}s_{j}\right)-\nonumber\\
 & -\left[\mbox{\ensuremath{\sum_{j\in I_{d}}s_{gj}}}-k_{g}\frac{1-\delta_{g}}{1-\delta_{g}\delta_{gE}}\right]^{+}\frac{1-\delta_{gE}}{1-\delta_{g}\delta_{gE}}\nonumber\\
 & -\min\{\mbox{\ensuremath{\sum_{j\in I_{d}}s_{gj}}},k_{g}\frac{1-\delta_{g}}{1-\delta_{g}\delta_{gE}}\}(1-\delta_{gE}).\nonumber\\
m_{i},s_{ij},k_{i},r_{i} & \geq0,\quad\forall i,j.\nonumber
\end{align}
\end{algorithmic}
\caption{Same as Algo 1 but with exact $E_i$ calculation in polynomial time.}
\label{algo3}
\end{algorithm}
}

\subsection{Exact calculation of $E_g$}
The LP in Algo 3 achieves a polynomial time calculation of $E_g$. 
As we mentioned in the paper, to do so, 
we need to assume that network nodes do an additional operation:
every node in the network uniformly at random mixes its incoming random packets before forwarding them towards Bob; we thus ensure that "all packets are treated equally". We then reduce the problem to calculating, what fraction of random packets that go through a given node, reach Bob.
Note that Alice needs to know the linear combinations of the random packets to be able to reproduce them when establishing the secret key with Bob.

Consider a directed acyclic graph, where there is an implicit partial ordering of edges. We say that edge $g < j$ if there exists a directed path that connects edge $g$ to $j$. 
The basic idea in the LP is to keep track of what amount, of  the random packets $s_g$ at edge $g$, are part to  the random packets in $s_j$,
with $g<j$.  
In the LP, the  variables $s_{gj}$ are used to denote random packets that  have passed through edge $g$ and also through
edge $j$, with $g<j$. For consistency of notation, we use $s_{gg}$ instead of $s_g$.  

We think of the $s_{gj}$ as "virtual flows", similarly to the approach in \cite{Li2006}.
Thus, we require that 
\[ \forall g,j \in \mathcal{E}, \mbox{with} \; \; g<j, \quad
s_{gj}  \leq s_{jj}
\]
Consider now a node $u$. The quantity
\[ A= \sum_{j \in I_{u}}s_{jj}-\sum_{j\in O_{u}}s_{jj} \]
captures how many of the random packets that are
incoming to node u, reach the "next hop" nodes towards Bob.
The quantity
\[
B= \sum_{j\in I_{u}}s_{gj}-\sum_{j\in O_{u}}s_{gj}\]
is the virtual flow in these packets that  has also passed through $g$.
We require in the LP that 
\[B\leq A.\]
Because intermediate nodes form and propagate linear combinations of packets,
we can let the LP  assign (consistently with the constraints) virtual flow values
that maximize the secure message rate.
We  calculate the part of $s_g$ that Bob received as $\sum_{j\in I_d}s_{gj}$.
Note that the last equation (that includes the $\min$) can  be easily written in linear form.

\subsection{Security for Algo 1}
As mentioned earlier, security follows directly by applying Theorems 10 and 11 of \cite{athan2014} as well as Lemma 4 of  \cite{JDFPA10}. For completeness we include  here a proof; this does not use the above but follows
the proof approach in \cite{athan2014}.

We denote by,

\[
b_{g}=m_{g}\frac{1-\delta_{gE}}{1-\delta_{g}\delta_{gE}}N
\]

\[
c_{g}=\left(\left(\sum_{j\in I_{D}}s_{j}\right)-\left(r_{g}+k_{g}(1-\delta_{g})\right)\frac{1-\delta_{gE}}{1-\delta_{g}\delta_{gE}}\right)N
\]

\[
+\varTheta(N^{3/4}).
\]

The $c_{g}$ are (with high probability) the number of secure packets that Bob has received
after $N$ time slots, given that: 1. All the packets that passed through edge $g$ actually
reached Bob (conservative assumption), 2. Edge $g$ was the one that was eavesdropped. This
concentration result is proved, using the Chernoff-Hoeffding bound, as follows,

\[
\Pr\{|C_{g}-c_{g}|\geq N^{3/4}|G_{E}=g\}=\exp\left(-\frac{\sqrt{N}}{4}\right)=o(N),
\]

where $C_{g}$ is the random variable of the number of secure packets that Bob has received
after $N$ time slots. Also, we use the random variable $G_{E}$ to denote the edge that is
eavesdropped.

It is,

\[
I(W;Z^{n}S^{n})=I(W;W_{I}),
\]

where with $W_{I}$ we denote the packets that are heard by Eve and $I$ is the set of indices
of these overheard columns. We know that,

\[
H(W_{I}|\left|I\right|=i,G_{E}=g)\leq i.
\]

Furthermore, from the MDS property of the $A$ matrix, we have,

\begin{eqnarray*}
H(W_{I}|W,\left|I\right|=i,G_{E}=g) & = & H(QA|W,\left|I\right|=i,G_{E}=g)\\
 & \geq & \min{\left\{ i,c_{g}\right\} }
\end{eqnarray*}

Thus,

\begin{eqnarray*}
 &  & I(W;Z^{n}S^{n})\\
 & = & H(W_{I})-H(W_{I}|W)\\
 & \leq & \sum_{g\in G}(\Pr\{G_{E}=g\}\sum_{i=0}^{N}\Pr\{|I|=i|G_{E}=g\}\\
 &  & (i-\min{\left\{ i,c_{g}\right\} }))\\
 & \leq & \sum_{g\in G}(\Pr\{G_{E}=g\}\sum_{i=0}^{N}\Pr\{|I|>c_{g}|G_{E}=g\}\\
 &  & (i-\min{\left\{ i,c_{g}\right\} }))
\end{eqnarray*}

We need a concentration result for $|I|$ by using the erasure channel probabilities. By inspecting
the ARQ scheme, the probability that a given encrypted message pakcet is received correctly
by Eve is, 
\[
p=(1-\delta_{gE})+\delta_{g}\delta_{gE}(1-\delta_{gE})+\dots=\frac{1-\delta_{gE}}{1-\delta_{g}\delta_{gE}}.
\]
Then, $|I|$ can be seen as a sum of $m_{g}N$ independent random variables on $\{0,1\}$ drawn
from a Bernoulli $Ber(p)$ distribution. So, we have that, 
\begin{align*}
E[|I|] & =m_{g}N\frac{1-\delta_{E}}{1-\delta\delta_{E}}=b_{g}.
\end{align*}

And, from the Chernoff-Hoeffding bound,

\[
\Pr\{|I|>c_{g}|G_{E}=g\}=\Pr\left\{ |I|\geq b_{g}+\left(c_{g}-b_{g}\right)\right\} \leq\exp\left(-\frac{\left(c_{g}-b_{g}\right)}{4}\right).
\]

Thus,

\begin{eqnarray*}
I(W;Z^{n}S^{n}) & \leq & \sum_{g\in G}(\Pr\{G_{E}=g\}\sum_{i=0}^{N}\exp\left(-\frac{\left(c_{g}-b_{g}\right)}{4}\right)\\
 &  & (i-\min{\left\{ i,c_{g}\right\} })),
\end{eqnarray*}

which goes to zero as $N$ grows.

{\small
\begin{algorithm}
\floatname{algorithm}{Algo 4}
\begin{algorithmic}
\renewcommand{\algorithmicrequire}{\textbf{Input:}}
 \REQUIRE Set of erasure probabilities $\delta_i$ and $\delta_{iE}$, number of eavesdropped edges V.
\renewcommand{\algorithmicrequire}{\textbf{Output:} Secure message rate and achievability scheme parameters}
  \REQUIRE 
\begin{align}
\max R,\text{s.t.:}\nonumber\\
R & =\sum_{i\in I_{D}}m_{i}\nonumber\\
\forall u\in \mathcal{V}-\{s,d\}:\nonumber\\
\sum_{i\in I_{u}}m_{i} & =\sum_{j\in O_{u}}m_{j}\nonumber\\
\sum_{i\in I_{u}}s_{i} & \geq\sum_{j\in O_{u}}(k_{j}+r_{j})\nonumber\\
\forall g\in \mathcal{E}:\nonumber\\
s_{g} & =r_{g}+k_{g}\frac{1-\delta_{g}}{1-\delta_{g}\delta_{gE}}\label{eq:delivered_randomness-3}\nonumber\\
1 & \geq\frac{r_{g}}{1-\delta_{g}}+\frac{k_{g}}{1-\delta_{g}\delta_{gE}}+\frac{m_{g}}{1-\delta_{g}}\nonumber\\
\forall G_{E}\underset{V}{\subset}\mathcal{E},\forall g\in G_{E}:\nonumber\\
m_{g}\frac{1-\delta_{gE}}{1-\delta_{g}\delta_{gE}} & \leq\left(\sum_{j\in I_{D}}s_{j}\right)-\sum_{g\in G_{E}}r_{g}\frac{1-\delta_{gE}}{1-\delta_{g}\delta_{gE}}\nonumber\\
 & \;\:-\sum_{g\in G_{E}}k_{g}\frac{(1-\delta_{gE})(1-\delta_{g})}{1-\delta_{g}\delta_{gE}}.\nonumber\\
m_{i},s_{i},k_{i},r_{i} & \geq0,\quad\forall i.\nonumber
\end{align}
\end{algorithmic}
\caption{Eve observing multiple edges}
\label{algo4}
\end{algorithm}
}

\subsection{Extensions of LPs}

We here provide  LPs that extend Algo 1 \& Algo 2, as discussed in Section~\ref{sec:ext}. 

\subsubsection{Algo 4: Eve observing multiple edges}
Algo 4 presents a case where Eve observes multiple ($V$) edges; what changes in this case is that, $E_i$ needs to account for all packets that Eve (and Bob) may have received, when Eve wiretaps edge $i$ and any other $V-1$ edges. 
Algo 4 is a variation of Algo 1; note that its complexity increases exponentially with the number of wiretapped edges $V$.

In particular, we follow the conservative assumption that all the eavesdropped packets reach Bob, and they
are all different to each other (which may not be since the same packet may be heard again
by Eve in a different edge). Thus, in the security constraint of the LP we subtract from the
total number of Alice-Bob shared packets, the number of packets that were heard by Eve in all
the channels she overhears,
\begin{eqnarray*}
m_{g}\frac{1-\delta_{gE}}{1-\delta_{g}\delta_{gE}} & \leq & \left(\sum_{j\in I_{D}}s_{j}\right)-\sum_{g\in G_{E}}r_{g}\frac{1-\delta_{gE}}{1-\delta_{g}\delta_{gE}}\\
 &  & \;\:-\sum_{g\in G_{E}}k_{g}\frac{(1-\delta_{gE})(1-\delta_{g})}{1-\delta_{g}\delta_{gE}}.
\end{eqnarray*}
The notation $G_{E}\underset{V}{\subset}\mathcal{E}$ denotes that $G_{E}$ is a subset of $\mathcal{E}$
with cardinality $V$.

\subsubsection{Algo 5: Multiple sources}
Algorithm Algo~5 presents the extension of Algo 2 in the case where there are $L$ sources, and each has a different message to send to a common destination. At a first glance, it might seem that the best we could do would simply be time-sharing between $L$ different secure message transmissions (from the $L$ sources to the receiver).
However, during the key-creation phase, \emph{we can}  exploit random packets originating from a given source, say source one, to create link-by-link keys that will be used to better protect a message send by say a source two. In particular, we can pull together the randomness generated by all sources to create a "universal" link-by-link key that protects all messages through that link. 

{\small
\begin{algorithm}
\floatname{algorithm}{Algo 5}
\begin{algorithmic}
\renewcommand{\algorithmicrequire}{\textbf{Input:}}
 \REQUIRE Set of erasure probabilities $\delta_i$ and $\delta_{iE}$.
\renewcommand{\algorithmicrequire}{\textbf{Output:} Secure message rate and achievability scheme parameters}
  \REQUIRE 
\begin{align}
\max\overset{L}{\underset{l=1}{\sum}}R_{l},\text{s.t.:}\nonumber\\
\forall l\in[1,L]:\nonumber\\
R_{l} & =\sum_{i\in I_{D}}m_{li}\nonumber\\
\forall u\in \mathcal{V}-\{s,d\},\forall l\in[1,L]:\nonumber\\
\sum_{i\in I_{u}}m_{li} & =\sum_{j\in O_{u}}m_{lj}\nonumber\\
\sum_{i\in I_{u}}s_{li} & \geq\sum_{j\in O_{u}}(k_{lj}+r_{lj})\nonumber\\
\forall g\in \mathcal{E}:\nonumber\\
1 & \geq\frac{\overset{L}{\underset{l=1}{\sum}}r_{lg}}{1-\delta_{g}}+\frac{\overset{L}{\underset{l=1}{\sum}}k_{lg}}{1-\delta_{g}\delta_{gE}}+\frac{\overset{L}{\underset{l=1}{\sum}}m_{lg}}{1-\delta_{g}}\nonumber\\
\overset{L}{\underset{l=1}{\sum}}w_{lg} & \leq\left(\overset{L}{\underset{l=1}{\sum}(}r_{lg}+k_{lg})\right)\frac{\delta_{gE}(1-\delta_{g})}{1-\delta_{g}\delta_{gE}}\nonumber\\
\forall g\in \mathcal{E},\forall l\in[1,L]:\nonumber\\
s_{lg} & =r_{lg}+k_{lg}\frac{1-\delta_{g}}{1-\delta_{g}\delta_{gE}}\nonumber\\
s_{lg} & =\sum_{p\in P:g\in p}s_{lp}\nonumber\\
m_{lg}\frac{1-\delta_{gE}}{1-\delta_{g}\delta_{gE}} & \leq w_{lg}+\sum_{p\in P_{-g}^{'}}s_{lp}\nonumber\\
m_{ij},s_{ij},k_{ij},r_{ij},w_{ij} & \geq0,\quad\forall i,j.\nonumber
\end{align}
\end{algorithmic}
\caption{LP with multiple sources located on different nodes}
\label{algo4}
\end{algorithm}
}

 We denote by $m_{li}$ the message rate at edge $i$ of the packets
of source $l$. We use the notation $s_{li}$, $r_{li}$, $k_{li}$.  We impose a time sharing (capacity) constraints at each edge, for the sum of the packets
that flow in that edge:
\begin{eqnarray*}
1 & \geq & \frac{\overset{L}{\underset{l=1}{\sum}}r_{lg}}{1-\delta_{g}}+\frac{\overset{L}{\underset{l=1}{\sum}}k_{lg}}{1-\delta_{g}\delta_{gE}}+\frac{\overset{L}{\underset{l=1}{\sum}}m_{lg}}{1-\delta_{g}}
\end{eqnarray*}
Each source functions independently, sending  random packets, creating the end-to-end
key and encrypting its message with it. This key is shared only between the specific sender and Bob. Thus
we cannot use it to encrypt end-to-end the messages of the other sources.
However, all the random packets (from
all sources) that flow through an edge can be used to create one universal link-by-link
key. This key can be used to encrypt all the packets (with link-by-link encryption), since
the key will be pealed of in the next node. Thus the size of the key is,
\[
\left(\overset{L}{\underset{l=1}{\sum}(}r_{lg}+k_{lg})\right)\frac{\delta_{gE}(1-\delta_{g})}{1-\delta_{g}\delta_{gE}}.
\]
We denote with $w_{lg}$ the amount will used in for the link-by-link encryption of the message
of transmitter $l$. Of course, the total amount of these parts cannot be bigger that the amount
of the universal key we created,
\begin{eqnarray*}
\overset{L}{\underset{l=1}{\sum}}w_{lg} & \leq & \left(\overset{L}{\underset{l=1}{\sum}(}r_{lg}+k_{lg})\right)\frac{\delta_{gE}(1-\delta_{g})}{1-\delta_{g}\delta_{gE}}
\end{eqnarray*}
In the security constraint, $w_{lg}$ takes the place of the link-by-link key in edge
$g$ for the message of transmitter $l$,

\begin{eqnarray*}
m_{lg}\frac{1-\delta_{gE}}{1-\delta_{g}\delta_{gE}} & \leq & w_{lg}+\sum_{p\in P_{-g}^{'}}s_{lp}.
\end{eqnarray*}

\end{document}